\documentclass[runningheads]{llncs}
\usepackage[T1]{fontenc}
\usepackage{comment}
\usepackage{graphicx}
\usepackage{soul}
\usepackage{color}
\begin{document}
\title{Personalized Travel Recommendations Using Wearable Data and AI: The INDIANA Platform}
\titlerunning{INDIANA: Personalized Travel Recommendations Using Wearables and AI}
%
\author{A. Manos\inst{1} \and
D.E. Filipidou\inst{1} \and
I. Deliyannis\inst{1} \and
N. Pavlidis\inst{1}\and
V. Argyros\inst{1} \and
I. Mazi\inst{1}
}

\authorrunning{A. Manos et al.}
\institute{Dotsoft SA \\
Poseidonos 71, Pilea, 55535, Thessaloniki, Greece\\
\email{info@dotsoft.gr}\\
\url{https://dotsoft.gr/} \and
Ionian University \\
Ioannou Theotoki 72, Kerkira 49100, Kerkira, Greece\\
\email{audiovisual@ionio.gr}\\
\url{https://ionio.gr/gr/}}
\maketitle              
\begin{abstract}
This work presents a personalized travel recommendation system developed as part of the INDIANA platform, designed to enhance the tourist experience through tailored activity suggestions, by leveraging data from wearable devices (e.g., Withings watches), user preferences, current location, weather forecasts, and activity history to provide real-time, context-aware recommendations. The platform is structured around three key use cases of tourism recommendations, including: (1) Daily itinerary planning based on user preferences, location, and weather conditions, (2) Immediate activity suggestions that consider the user’s current situation and environment, (3) Automatic, proactive recommendations triggered by user activity and health data. The platform not only supports individual tourists in maximizing their travel experience but also offers insights to tourism professionals to enhance service delivery, and by integrating modern technologies such as AI, IoT, and wearable analytics, it provides a seamless, personalized, and engaging experience for travelers.

\end{abstract}
\section{Introduction}\label{sec:Intro}

The demand for personalized travel experiences has grown significantly as modern travelers seek recommendations that align with their unique preferences, current location, and real-time context. As travelers increasingly prioritize customization in their travel journeys, there is a parallel rise in platforms that leverage artificial intelligence (AI) and Internet of Things (IoT) technologies to offer tailored suggestions. Wearable devices and smartphones provide constant streams of data that can reveal users’ preferences and situational needs, such as current health metrics or environmental conditions. This emerging capacity for real-time data analysis allows platforms to adapt their recommendations based on users' physical states and surroundings, enhancing the travel experience dynamically~\cite{Zheng2011Learning}.
However, despite these technological advancements, most travel recommendation systems have limitations. Many platforms rely primarily on static data or infrequent updates, which restricts their ability to respond to real-time changes in user context. For example, some systems focus on travel history to suggest destinations, which is effective for planning but lacks adaptability for on-the-go adjustments~\cite{Majid2013A}. Additionally, social media and crowdsourced platforms use user-generated content, such as geotagged photos, to provide recommendations based on general trends rather than individual, real-time factors~\cite{Majid2013A}. These systems lack the integration of user health and environmental data, which can be critical in providing more relevant and context-aware suggestions.
The INDIANA platform addresses these gaps by leveraging real-time data from multiple sources—wearable devices, location services, weather APIs, and user input—to deliver a dynamic, highly personalized recommendation experience. Unlike static or social media-driven recommendation systems, INDIANA’s AI-powered engine integrates user-specific health data and environmental factors to continuously refine recommendations based on changing user needs and conditions. This capability positions the INDIANA platform as a pioneering solution in the field of adaptive travel recommendations, responding to each traveler’s unique and evolving context in real time.

\section{State of the Art}

Artificial Intelligence (AI) has revolutionized numerous aspects of daily life, transforming industries such as telecommunications~\cite{pavlidis2023intelligent}, healthcare~\cite{briola2024federated}, and education~\cite{pavlidis2024federated}.
Consequently, the field of personalized travel recommendation has been shaped by several innovations in AI, IoT, and data-driven analytics. Various systems have emerged to tackle the demand for customized travel experiences, yet significant limitations persist in many existing solutions.

Early travel recommendation systems are largely based on static datasets that analyze user history or general travel trends to offer suggestions. For instance, many platforms use travel histories and general user profiles to suggest similar destinations but do not adapt these suggestions in real time~\cite{Zheng2011Learning}. While effective for broad planning, these systems often fall short in delivering context-sensitive recommendations that respond to users' immediate circumstances, such as location or current weather conditions.

A second class of systems leverages social media data and crowdsourced content, such as location-tagged images and reviews, to build recommendations based on popular attractions or user reviews~\cite{Majid2013A}. These systems, including platforms that mine data from sources like Flickr and Twitter, provide valuable insights based on community trends but lack the precision of personalized recommendations. For instance, users receive suggestions based on general trends or collective data, which may not align with their specific needs, preferences, or current physical conditions.

Recent advances have introduced systems that incorporate IoT and wearable data, aiming to tailor recommendations based on a user’s physical condition or environment. Liu et al. (2015)~\cite{Liu2015A} presented a smart tourism recommendation platform that uses IoT to provide route suggestions based on location and preference, optimizing itineraries for cost and efficiency. However, these systems often lack the ability to update recommendations in real time based on continuous health or environmental inputs. Similarly, systems like the IoT-driven tour guide platforms discussed by Arvaniti-Bofili et al. (2020)~\cite{arvaniti2020towards} use historical location data and clustering to deliver suggestions, but their reliance on past data limits adaptability for dynamic, real-time changes.

Some platforms, such as those incorporating real-time GPS and feedback loops, represent a step forward by updating recommendations as users move through locations. Systems like Hu et al. (2018)~\cite{Hu2018Personalized} utilize social media data and GPS tracking to produce context-sensitive suggestions but are limited by the scope of data integration, primarily relying on location without incorporating user health metrics or physical conditions.

The INDIANA platform advances beyond these limitations by integrating multiple data sources, including real-time health metrics, location, weather, and user preferences, into a single, adaptive recommendation engine. Unlike static or history-based systems, INDIANA’s machine learning engine adjusts recommendations dynamically, factoring in the traveler’s current location, physical state, and immediate context, such as nearby attractions and real-time weather conditions. This comprehensive approach allows INDIANA to deliver a uniquely responsive and personalized travel experience, continuously refining its suggestions as users’ needs and circumstances evolve.

In summary, while many existing systems provide valuable travel recommendations, they are often limited by static data, historical trends, or single-source information. INDIANA’s multi-source, real-time integration represents a significant innovation, setting a new standard in personalized, context-aware travel recommendation systems. This platform not only enhances user engagement but also supports users in making travel decisions that align with their preferences and physical well-being, demonstrating the full potential of AI and IoT in the personalized travel sector.

\section{Proposed Approach - Indiana platform}

The INDIANA platform is at this point a conceptual design aim to transform the tourist experience by providing smart, personalized recommendations tailored to individual preferences and real-time conditions, with its primary goal being to enhance the way travelers interact with their surroundings, ensuring that every experience is aligned with their interests and current context \cite{si2024intelligent}. Through the integration of advanced technologies, the platform processes real-time data by leveraging human-centred AI to offer recommendations that are not only relevant but also timely, thereby enriching the overall travel experience. It supports a variety of use cases, demonstrating its versatility in meeting the diverse needs of travelers, including:

\begin{itemize}
\item \textbf{Daily Itineraries Generation} with the platform generating comprehensive daily plans based on user preferences, location, and weather forecasts, ensuring that travelers can maximize their time and enjoyment at each destination.

\item \textbf{Instant Recommendations} with the user interacting with INDIANA platform using chatbot to obtain instant recommendations for activities based on his/her current location and circumstances.

\item \textbf{Pop-Up Recommendations} by leveraging data from wearables and mobile devices, the platform can autonomously suggest activities without user prompting, facilitating a proactive approach to travel planning.
\end{itemize}

By focusing on these use cases, INDIANA not only enhances user engagement but also ensures that travelers can enjoy unique and meaningful experiences that align with their individual preferences.
The primary objectives of this paper are twofold. Firstly, this study aims to evaluate the role of wearable technologies and artificial intelligence in enhancing the personalization of travel recommendations, by analyzing data collected from devices such as smartwatches and mobile applications, it assess how these technologies can provide real-time, context-aware suggestions that align with individual preferences and situational factors. 
Secondly, this paper will also highlight the innovative applications of real-time data and smart devices in transforming the tourism sector. By examining case studies and specific use cases supported by the INDIANA platform, the research will emphasize the potential for smart technologies to create more engaging and personalized travel experiences and at the same time seeking to advance the understanding of how integrated technological solutions can redefine tourism through enhanced personalization and user engagement~\cite{sousa2024use}.

\section{Methodology}\label{sec2}

\subsection{System Architecture}
\begin{figure}[h]
    \centering
    \includegraphics[width=0.75\textwidth]{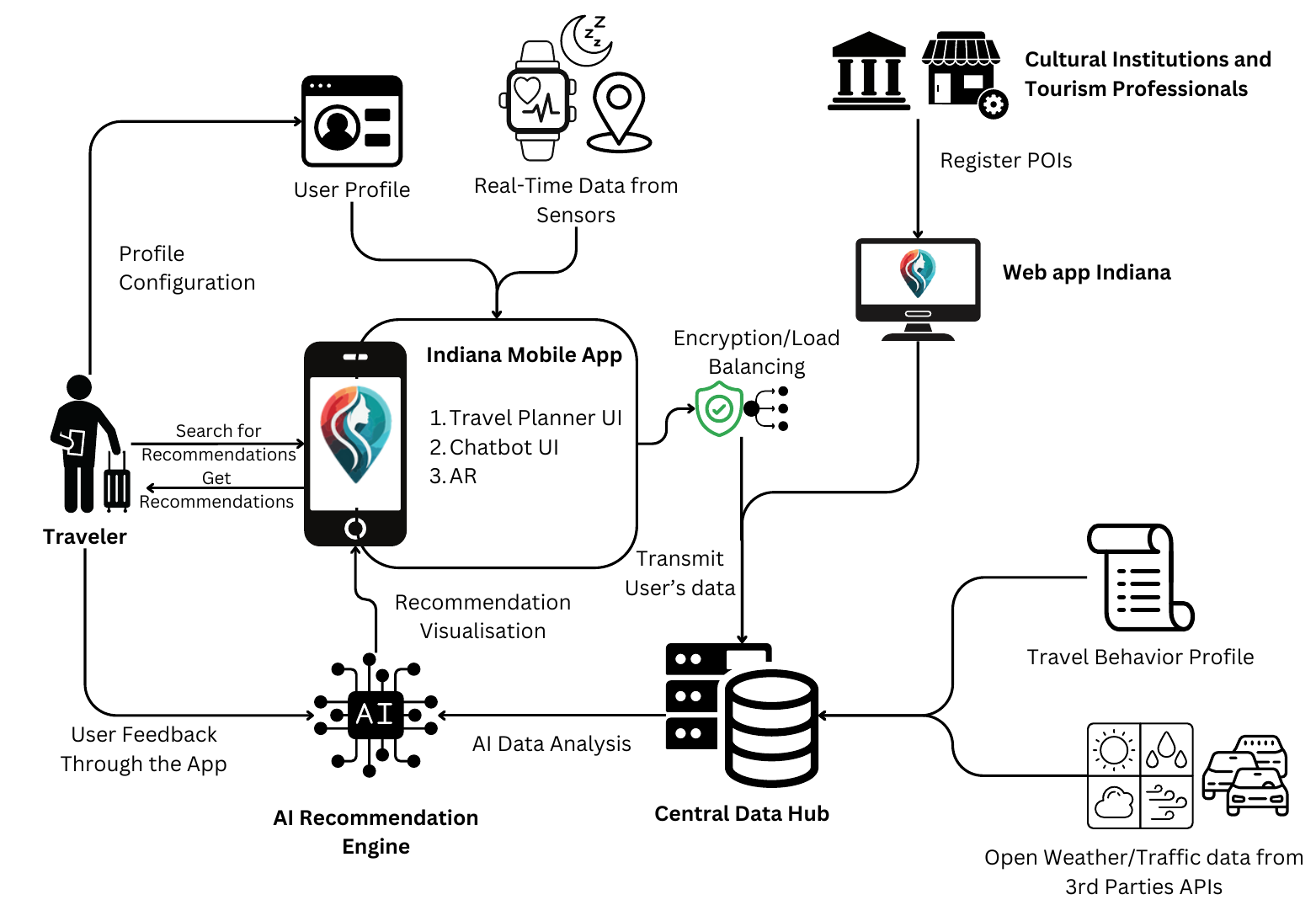}
    \caption{Indiana Architecture}
    \label{fig:architecture}
\end{figure}

The INDIANA platform integrates multiple data streams and processing components to deliver real-time, personalized travel recommendations based on user preferences, health data, and environmental context. The architecture can be divided into several layers, each responsible for handling different aspects of the system's functionality. A comprehensive overview of platform's architecture is presented in Fig. \ref{fig:architecture}.

\begin{enumerate}
    \item \textbf{Data Collection Layer}: Smartphones are used to collect location data using the Fused Location API or GPS. The platform also syncs weather data through APIs, such as OpenWeather or similar services, ensuring that recommendations are adjusted according to real-time environmental conditions. Additionally, users can manually input preferences (e.g., interests in cultural or gastronomic experiences) via the mobile app, providing an extra layer of customization with devices such as Withings smartwatches, Fitbit, and Android bands being used to collect health and activity data, including steps, heart rate, and sleep quality, which is then transmitted to the platform through APIs like Health Connect or Google Fit, which aggregate data from multiple devices for more comprehensive user profiles.
    \item \textbf{Data Processing Layer}: After collection, wearable data are processed in real time, monitoring user conditions (e.g., activity level, heart rate), allowing the platform to generate recommendations that are suitable for the user's physical state at a given moment with the raw health data being anonymized to ensure privacy, and only necessary data for recommendation generation are retained, with location data from mobile devices being paired with contextual information such as weather and local events. The INDIANA platform accesses APIs to pull in up-to-date weather forecasts and events happening nearby, allowing the system to generate location-aware recommendations that account for external conditions like weather, time of day, and proximity to attractions.
    \item \textbf{Recommendation Engine}: The platform's core recommendation engine uses machine learning algorithms to tailor suggestions based on individual user profiles, built by analyzing health metrics, activity data, and manually input preferences. The engine dynamically adjusts recommendations as new data comes in, such as a change in heart rate or a shift in location, ensuring that suggestions are always relevant to the user's current state and context. By incorporating both user-specific data and contextual information (e.g., weather conditions, events), the platform can provide highly relevant suggestions, recommending indoor activities when the weather is poor, or suggesting low-intensity experiences when the user's heart rate is elevated, making the system adaptable and responsive.
    \item \textbf{User Interface and Feedback Loop}: The mobile application serves as the main point of interaction for users, where they can receive recommendations, input preferences, and provide feedback on their experiences, improving the recommendation engine over time, allowing the system to refine its suggestions based on user satisfaction and preferences. After users engage with the platform’s recommendations, they are prompted to complete surveys through email or in-app notifications, with the collected data providing quantitative insights into user satisfaction and serve as a basis for continuous system improvement.
    \item \textbf{Data Privacy and Security}: To comply with data privacy regulations, all collected raw data (health, activity, location) are anonymized immediately after processing with the platform retains only the information required for generating recommendations, and all raw data are deleted from the servers after use, ensuring that user privacy is maintained while still delivering personalized experiences.
\end{enumerate}

\subsection{Data Collection Methods}
Data collection for the INDIANA platform will be conducted through multiple channels to ensure a comprehensive understanding of user experiences and the effectiveness of the personalized recommendations. Data from wearable devices, specifically Withings smartwatches, will be collected to monitor user health metrics and location data in real-time, providing valuable insights into users' physical conditions and activity levels, allowing for personalized recommendations \cite{kim2024artificial} with the smartwatches tracking various health metrics, including sleeping time, heart rate, steps taken, and overall activity levels, transmitting this data securely to the INDIANA platform to ensure user privacy and data integrity, with GPS functionality enabling real-time location tracking, allowing the platform to tailor recommendations based on users’ current positions, ensuring that suggestions were contextually relevant by taking into account proximity to points of interest, facilitating dynamic adjustments to recommendations, responding to changes in user health status or activity levels throughout the day.

To enhance the relevance and accuracy of the recommendations, contextual data such as weather forecasts and local event information will be incorporated, with real-time weather data being accessed through Application Programming Interfaces (APIs) from reputable weather services, including information about temperature, precipitation forecasts, and current conditions, allowing the platform to suggest activities that aligned with the weather \cite{tussyadiah2012role}. Additionally, information about local events, festivals, and activities will be sourced through event collection databases, helping the platform recommend timely and engaging activities based on user preferences, while also ensuring that all contextual data remained accurate and up-to-date, thereby improving the reliability of recommendations and enhancing user trust in the platform.

\section{Discussion}\label{sec: Discussion}

The INDIANA platform represents a significant advancement in personalized travel experiences, leveraging wearable technology and artificial intelligence to deliver context-aware recommendations, with the anticipated outcomes suggesting that integrating real-time health and location data from wearables will enhance the user experience, enabling more tailored suggestions that align with individual preferences and circumstances~\cite{buhalis2015smart}, revealing insights into user satisfaction, particularly regarding the relevance and timeliness of the recommendations provided with previous studies indicating that personalization enhances user engagement, and the results from INDIANA possibly corroborating this finding~\cite{xiang2015can}. It is anticipated that users will find value in the platform's ability to suggest activities that align with both their personal interests and current physical conditions, as well as the surrounding environmental context with the inclusion of contextual data, such as weather forecasts and local events, improving the relevance and accuracy of recommendations, potentially leading to greater user satisfaction and a higher likelihood of engagement with the suggested activities.

However, the INDIANA platform has several important challenges. Protecting user privacy is critical, especially given the sensitive health and location data it uses. Frequent notifications may lead to user fatigue, so managing notification frequency is essential to keep users engaged. Additionally, the platform’s reliance on real-time data requires strong processing capabilities, which can be a limitation in areas with weak connectivity. Finally, ethical considerations around health data suggest that recommendations should support user decisions without being overly directive. Addressing these challenges is key to ensuring the platform’s success and user trust.

\section{Conclusion}\label{sec: Conclusion}

The INDIANA platform aims to provide personalized recommendations that significantly enhance the travel experience by harnessing real-time data on user health and location, the platform, underscoring the growing demand for tailored travel solutions that cater to individual preferences and situational contexts. The anticipated outcomes highlight the effectiveness of using wearable data to generate relevant and timely suggestions, contributing to higher user satisfaction and engagement. Additionally, the incorporation of contextual information, such as weather conditions and local events, is expected to further refine the personalization process, ensuring that recommendations align with the dynamic nature of travel environments. This research also contributes valuable insights into the potential of AI-driven, context-aware travel solutions, with the findings contributing to the ongoing development efforts and future studies aimed at optimizing user experiences in tourism.

\subsubsection{Acknowledgements} This research has been co‐financed by the European Regional Development Fund of the European Union and Greek national funds through the Operational Program Competitiveness, Entrepreneurship and Innovation, under the call "Partnerships between Enterprises and Research and Knowledge Transfer Organizations in the fields of RIS3 of the Region Ionian Islands with the id: IONP2-0075453.

\bibliographystyle{splncs04}
\bibliography{main}

\end{document}